\documentclass[a4paper]{article}

\usepackage{graphicx}
\usepackage{amsfonts}
\usepackage{amsmath}
\usepackage[a4paper]{geometry}
\usepackage{layouts}
\usepackage{float}
\usepackage{authblk}

\title{Natural statistics of binaural sounds}
\author[1]{Wiktor M\l ynarski\thanks{Corresponding author. Email: mlynar@mis.mpg.de}}
\author[1,2]{J\" urgen Jost}
\affil[1]{Max-Planck Institute for Mathematics in the Sciences, Leipzig, Germany}
\affil[2]{Santa Fe Institute, Santa Fe, New Mexico, USA }

\begin{document}
\maketitle

\begin{abstract}
Binaural sound localization is usually considered a discrimination task, 
where interaural time (ITD) and level (ILD) disparities at pure frequency 
channels are utilized to identify a position of a sound source. In natural 
conditions binaural circuits are exposed to a stimulation by sound 
waves originating from multiple, often moving and overlapping 
sources. Therefore statistics of binaural cues depend on acoustic 
properties and the spatial configuration of the environment. In order to process binaural sounds efficiently, the auditory system should be adapted to naturally encountered cue distributions. Statistics of cues encountered naturally and their dependence on the physical properties of an auditory scene have not been studied before. Here, we performed binaural recordings of three auditory scenes with varying spatial properties. We have analyzed empirical cue distributions from each scene by fitting them with parametric probability density functions which allowed for an easy comparison of different scenes. Higher order statistics of binaural waveforms were analyzed by performing Independent Component Analysis (ICA) and studying properties of learned basis functions.
Obtained results can be related to known neuronal mechanisms and suggest how binaural hearing can be understood in terms of adaptation to the natural signal statistics.
\end{abstract}

\section*{Introduction}

The idea that sensory systems reflect the statistical structure of stimuli encountered by organisms in their ecological niches \cite{Barlow, Attneave, RiekeSpikes} has driven numerous theoretical and experimental studies. Obtained results suggest that tuning properties of sensory neurons match regularities present in natural stimuli \cite{OlshausenSimoncelli}. In light of this theory, neural representations, coding mechanisms and anatomical structures could be undestood by studying characteristics of the sensory environment.

To date, natural scene statistics research have been focusing mostly
on visual stimuli \cite{HyvarinenBook}. Nevertheless, a number of
interesting results relating natural sound statistics to the auditory
system have also been delivered. For instance, Rieke et al
demonstrated that auditory neurons in the frog increase information
transmission, when the spectrum of the white-noise stimulus is shaped
to match the spectrum of a frog call \cite{RiekeFrog}. In a more
recent experiment, Hsu and colleagues \cite{TheunissenNeurons} have
shown similar facilitation effects in the zebra finch auditory system
using stimuli with power and phase modulation spectrum of a
conspecific song. In a statistical study it has been shown that
modulation spectra of natural sounds display a characteristic
statistical signature \cite{TheunissenMod} which allowed to form
quantitative predictions about neural representations and coding of
sounds. Other statistical models of natural auditory scenes have also
led to interesting observations. Low-order, marginal statistics of
amplitude envelopes, for instance, seem to be preserved across
frequency channels as shown by Attias and Schreiner
\cite{AttiasSchreiner}. This means that all locations along the
cochlea may be exposed to (on average) similar stimulation patterns in
the natural environment. A strong evidence of adaptation of the early
auditory system to natural sounds was provided by two complementary
studies by Lewicki \cite{Lewicki} and Smith and Lewicki
\cite{SmithLewicki}. The authors modeled high order statistics of
natural stimuli by learning sparse representations of short sound
chunks. In such a way, they reproduced filter shapes of the cat's
cochlear nerve. These results were recently extended by Carlson et al
\cite{Carlson} who obtained features resembling spectro-temporal
receptive fields in the cat's Inferior Colliculus by learning sparse
codes of speech spectrograms. Human perceptual capabilities have also
been related to natural sound statistics in a recent study by
McDermott and Simoncelli \cite{McDermottSimoncelli}. In a series of
psychophysical experiments the authors have shown that perceived
realism and recognizability of sound ''textures'' by human subjects
depends on how well the time-averaged statistics of stimulus
modulation correspond to those of natural sounds. The aquired body of evidence strongly suggests that neural representations of acoustic stimuli reflect structures present in the natural auditory environment.

The above mentioned studies investigated statistical properties of
single channel, monaural sounds relating them to the functioning of
the nervous system. However, in natural hearing conditions the sensory
input is determined by many additional factors - not only properties
of the sound source. Air pressure waveforms reaching the cochlea are
affected by positions and motion patterns of sound sources as well as
head movements of the listening subject. These spatial aspects generate differences between stimuli present in each ear ,
which are traditionally divided into two classes: interaural level and
phase differences \cite{GrotheMcAlpine}. The sound wavefront reaches
firstly the ipsilateral ear and after a very short time delay the contralateral one. This generates the interaural time difference (ITD). After cochlear filtering - in pure frequency channels, ITDs correspond to phase differences (IPDs). Additionally, sound received by the contralateral ear is attenuated by the head, which generates the interaural level difference (ILD). According to the widely acknowdledged duplex theory \cite{Rayleigh, GrotheMcAlpine}, in mammals, IPDs are used to localize low frequency sounds. The theory predicts that in higher frequency regimes IPDs become ambiguous and therefore sounds of frequency above a certain threshold (around $1.5$ kHz in humans) are localized based on ILDs which become more pronounced due to the low-pass filtering properties of the head. Binaural cues are of a relative nature and positions of auditory objects are not represented on the sensory epiphelium - the cochlear membrane - in a direct way. They are reflected in binaural cue values, which themselves vary with changing spatial configuration of the environment and depend on sound sources' spectra. 

Binaural hearing mechanisms have also been studied in terms of adaptation to natural stimulus statistics. Harper and McAlpine \cite{HarperMcAlpine} have shown that tuning properties of IPD sensitive neurons in a number of species can be predicted from distributions of this cue naturally encountered by the organism. This was done by forming a model neuronal representation of maximal sensitivity to the stimulus change, as quantified by the Fisher information. Two recent experimental studies revealed rapid adaptation of binaural neurons and perceptual mechanisms to changing cue statistics. Dahmen and colleagues \cite{Dahmen} stimulated human and animal subjects with non-stationary ILD sequences. They collected electrophysiological and psychophysical evidence in favor of adaptation to the stimulus distribution. Maier et al \cite{Maier}, in turn, have shown that neural tuning curves in the guinea pig and human performance in a localization task can be adapted to varying ITD distributions. Both - neural representation and human performance were, however, constrained to represent midline locations with the highest accuracy. One has to note that Maier et al, take an issue with the interpretation of results obtained by Dahmen et al. suggesting that they may be explained by adaptation to the sound level and not ILDs per se.

Adaptation of the binaural auditory system to changes in the cue
distribution occuring on different timescales seems to be
experimentally confirmed. Despite this fact, the statistical structure
of binaural sounds encountered in the natural environment and its
dependence on the auditory scene have not yet been
studied. In this paper we address this shortage. We performed binaural
recordings of three real-world auditory scenes characterized by
different acoustic properties and spatial dynamics. In the next step
we extracted binaural cues - IPDs and ILDs and studied their marginal
distributions by means of fitting parametric probability density
functions. Parameters of fitted distributions allowed for an easy
comparison of different scenes, and revealed which aspects change and which seem to
remain invariant in different auditory environments. To analyze
high-order statistics of binaural waveforms we performed Independent
Component Analysis (ICA) of the signal, and studied properties of the
learned features. The results obtained suggest how mechanisms of binaural hearing can be understood in terms of adaptation to natural stimulus statistics. They also allow for experimental predictions regarding neural computation and representation of the auditory space.

\section*{Results}

\subsection*{Binaural spectra}

In the first step of the analysis, monaural Fourier spectra were compared with each other. Frequency spectra of recorded sounds are displayed on figure \ref{fig:spectra}. Strong differences across all recorded auditory scenes were present. In two of them - the forrest walk scene and the city center scene, frequency spectrum had an exponential (power-law) shape, which is a characteristic signature of natural sounds \cite{Voss}. Since the nocturnal nature scene was dominated by grasshoper sounds, its spectrum had two dominant peaks around $7$ and $10$ kHz. 
In all three cases, sounds in both ears contained a similar amount of energy in lower frequencies (below $4$ kHz) - which is reflected by a good overlap of monaural spectra on the plots. In higher frequencies though, the spectral power was not always equally distributed in both ears. This difference is most strongly visible in the spectrum of the nocturnal nature scene. There, due to a persistent presence of a sound source (a grasshoper) closer to the right ear, corresponding frequencies were amplified with respect to the contralateral ear. Since the spatial configuration of the scene was static, this effect was not averaged out in time. Monaural spectra of the forrest walk scene overlapped to a much higher degree. A small notch in the left ear spectrum is visible around $6$ kHz. This is most probably due to the fact that the recording subject stood next to a stream flowing at his right side for a period of time. The city center scene, has almost identical monaural spectra. This is a reflection of its rapidly changing spatial configuration - sound sources of similar quality (mostly human speakers) were present in all positions during the time of the recording.    

\subsection*{Interaural level difference statistics}
\label{sec:ILD}
An example joint amplitude distribution in the left and the right ear
is depicted in figure \ref{fig:ILD1} A. It is not easily described by
any parametric probability density function (pdf), however monaural
amplitudes reveal a strong linear correlation. The correlation
coefficient can be therefore used as a simple measure of interaural
redundancy by indicating how similar the amplitude signal in both ears
is, at a particular frequency channel. High correlation values would
imply that both ears receive similar information, while low
correlations indicate that the signal at both sides of the head is
generated by different sources. Interaural amplitude correlations for
all recorded scenes are plotted as a function of frequency on figure
\ref{fig:ILD1} B. A general trend across the scenes is that
correlations among low frequency channels (below $1$ kHz) are strong
(larger than $0.5$) and decay with a frequency increase. Such a trend is expected due to the filtering properties of the head, which attenuates low frequencies much less than higher ones. The spatial structure of the scene also finds reflection in binaural correlation - for instance, a peak is visible in the nocturnal nature scene at $7$ kHz. This is due to a presence of a spatially fixed source generating sound at this frequency (see figure \ref{fig:spectra}). The most dynamic scene - city center - reveals, as expected, lowest correlations across most of the spectrum.

Interaural level differences ILD were computed separately in each frequency channel. Figure \ref{fig:ILD1} C displays an example ILD distribution (black line) together with a best fitting Gaussian (blue dotted line) and logistic distribution (red dashed line). Logistic distributions provided the best fit to ILD distributions for all frequencies and recorded scenes, as confirmed by the KS-test (results not shown). ILD distribution at frequency $\omega$ was therefore defined as
\begin{equation}
\label{ildPdf}
p(ILD_\omega|\mu_\omega,\sigma_\omega)=\frac{\exp(-\frac{ILD_\omega-\mu_\omega}{\sigma_\omega})}{\sigma_\omega(1+\exp(-\frac{ILD_\omega-\mu_\omega}{\sigma_\omega}))^2}
\end{equation}
where $\mu_\omega$ and $\sigma_\omega$ are frequency specific mean and
scale parameters of the logistic pdf respectively. The variance of the logistic distribution is fully determined by the scale parameter.

Empirical ILD distributions are plotted in figure \ref{fig:ILD2} A. As
can be immediately observed, they preserve similar shape in all
frequency channels and auditory scenes, regardless of their type. The
mean ($\mu_\omega$) and scale ($\sigma_\omega$) parameters of the
fitted distributions are plotted as a function of frequency in figures
\ref{fig:ILD2} B and C respectively. The mean of all distributions is
very close to $0$ dB in most cases. In the two non-static scenes, i.e., forrest walk and city center, deviations from $0$ are very small. Marginal ILD distributions of the spatially constant scene - nocturnal nature - were slightly shifted away from zero for frequencies generated by a sound source of a fixed position.
The scale parameter behaved differently than the mean. In all auditory scenes it grew monotonically with the increasing frequency. The increase was quite rapid for frequencies below $1$ kHz - from $1.5$ to $2$. For higher frequencies the change was much smaller and in the $1 - 11$ kHz interval $\sigma$ did not exceed the value of $2.5$.
What may be a surprising observation is the relatively small change in the 
ILD distribution, when comparing high and low frequencies. It is known
that level differences become much more pronounced in the high
frequency channels \cite{TheShapeOfEars}, and one could expect a
strong difference with a frequency increase. These results can be partially explained by observing a close relationship between Fourier
spectra of binaural sounds and means of ILD distributions. In a
typical, natural setting sound sources on the left side of the head
are qualitatively (spectrally) similar to the ones on the other side,
therefore the  spectral power in the same frequency bands remains similar in both ears. Average ILDs deviate from $0$ if a sound source was present at a fixed position during the averaged time period. Increase in the ILD variance (defined by the scale parameter $\sigma$) with increasing frequency, can be explained by the filtering properties of the head. While for lower frequencies a range of possible ILDs is low, since large spatial displacements generate weak ILD differences, in higher frequency regimes ILDs become more sensitive to the sound source position hence their variability grows. On the other hand, objects on both sides of the head reveal similar motion patterns and in this way reduce the ILD variability, which may account for the the small rate of change.
Despite observed differences, ILD distributions revealed a strong invariance to frequency and were homogenous across different auditory scenes.

\subsection*{Interaural phase difference statistics}

Marginal distributions of a univariate, monaural phases over a long time period are all uniform, since phase visits cyclically all values on a unit circle. An interesting structure appears in a joint distribution of left and right ear phase values from the same frequency channel (an example is plotted in figure \ref{fig:IPD1}). Monaural phases reveal dependence in their difference. This means that their joint probability is determined by the probability of their difference:
\begin{equation}
\label{eq:IPD1}
p(\phi_L, \phi_R) \propto p(\phi_L - \phi_R)
\end{equation}
where $\phi_L$ and $\phi_R$ are instantenous phase values in the left and the right ear respectively. The well known physical mechanisms explain this effect. The sound wavefront reaches first the ear ipsilateral to the sound source and then, after a short delay the contralateral one. The temporal difference generates a phase offset, which is reflected in the joint distribution of monaural phases. This simple observation implies, however, that IPDs constitute an intrinsic statistical structure of the natural binaural signal.

IPD histograms were well approximated by the von Mises distribution (additional structure was present in IPDs from the forrest walk scene - see subsection \ref{vmMixtures}). A distribution of two monaural phase variables revealing dependence in the difference can be then written as a von Mises distribution of their differences:
\begin{equation}
\label{eq:IPD2}
p(\phi_{L,\omega}, \phi_{R,\omega}) = p(IPD_\omega | \kappa_\omega, \mu_\omega) = \frac{1}{2\pi I_0(\kappa)}e^{\kappa \cos(IPD_\omega - \mu_\omega)}
\end{equation}
where $IPD_\omega = \phi_{L,\omega} - \phi_{R,\omega}$ is the IPD at frequency $\omega$, $\mu_\omega$ and $\kappa_\omega$ are frequency specific mean and concentration parameters and $I_0$ is the modified Bessel function of order $0$. In such a case, the concentration parameter  $\kappa$ controls mutual dependence of monaural phases \cite{CadieuPhase}. For large $\kappa_\omega$ values $\phi_{L,\omega}$ and $\phi_{R,\omega}$ are strongly dependent and the dependence vanishes for $\kappa = 0$.

\subsubsection*{IPD distributions}

Figure \ref{fig:IPD2} A depicts IPD histograms in all scenes depending on the frequency channel. Thick black lines mark $IPD_{\omega, max}$ - the ''maximal IPD'' value i.e. phase displacement corresponding to a time interval required for a sound to travel the entire interaural distance. $IPD_{\omega, max}$ can be computed in a following way. Assuming a spherical head shape, the time period required by the sound wave to travel the distance between the ears is equal to:
\begin{equation}
\label{ITD}
ITD = \frac{R_{head}}{v_{snd}} (\Theta + \sin(\Theta))
\end{equation}
where $R_{head}$ is the head radius, $v_{snd}$ the speed of sound and $\Theta$ the angular position of the sound source measured in radians from the midline. The ITD is maximized for sounds located directly oposite to one of the ears, deviating from the midline by $\frac{\pi}{2}$ ($\Theta = \frac{\pi}{2}$). $ITD_{max}$ becomes 
\begin{equation}
\label{ITDmax}
ITD_{max} = \frac{R_{head}}{v_{snd}} (\frac{\pi}{2} + 1).
\end{equation}
The maximal IPD is then computed separately in each frequency channel $\omega$
\begin{equation}
\label{IPDmax}
IPD_{\omega, max} =  2 \pi \omega ITD_{max}.
\end{equation}
The above calculations assume a spherical head shape, which is a major
simplification. It is, however, sufficient for the sake of the current analysis.

At low frequencies most IPD values do not exceed the ''forbidden''
line, and the resulting plot has a triangular shape. This is a common tendency in IPD distributions, visible across all auditory scenes. Additionally, due to phase wrapping, for frequencies where $\pi \le |IPD_{max}| \le 2\pi$ the probability mass is shifted away from the center of the unit circle towards the $-\pi$ and $\pi$ values, which is visible as blue, circular regions in the middle of the plot. This trend is not present in the forrest walk scene, where a clear peak at $0$ radians is visible for almost all frequencies. This figure can be compared with figure 3 in \cite{HarperMcAlpine} and 14 in \cite{GoodmanBrette}.
The two panels below, i.e., figures \ref{fig:IPD2} B and C, display
plots of the $\kappa$ and $\mu$ parameters of von Mises distributions
as a function of frequency. The concentration parameter $\kappa$
decreases in all three scenes from a value close to $1.5$ (strong concentration) to below $0.5$ in the $200$ Hz to $500$ Hz interval, which seems to be a robust property in all environments. Afterwards, small rebounds are visible. For auditory scenes recorded by a static subject, i.e., nocturnal nature and city center, rebounds occur at frequencies where $IPD_{max}$ corresponds to $\pi$ multiplicities (this is again, an effect of phase wrapping). The $\kappa$ value is higher for a more static scene - nocturnal nature - reflecting a lower IPD variance. For frequencies above $2$ kHz, concentration converges to $0$ in all three scenes. This means that IPD distributions become uniform and monaural phases mutually independent.
The frequency dependence of the position parameter $\mu$ is visible on figure
\ref{fig:IPD2} C. For the forrest walk scene, IPD distributions were centered
at the $0$ value with an exception at $700$ Hz. For the two scenes recorded by a static subject
, distribution peaks were roughly aligned along the $IPD_{max}$ as
long as it did not exceed $-\pi$ or $\pi$ value. In higher frequencies
they varied much stronger, although one has to note that for
distributions close to uniform ($\kappa \rightarrow 0$), position of the peak becomes an ill defined and arbitrary parameter.

Equations \ref{ITD} - \ref{IPDmax} allow to compute the ''maximal''
IPD value ($IPD_{max}$), constrained by the size of the organism's
head. A single, point sound source in an anechoic environment would
never generate IPD exceeding $IPD_{max}$. In natural hearing
conditions however, such IPDs occur due to the presence of two sound
sources at both sides of the head or due to acoustic reflections
\cite{GrotheMcAlpine}. Their presence is visible in figure
\ref{fig:IPD2} as  probability mass lying outside of the black lines
marking maximal IPD values at particular frequencies. Figure
\ref{fig:IPD3} displays a proportion of IPDs larger than the one
defined by the head size plotted against frequency. The lines
corresponding to three recorded auditory environments lie in parallel
to each other, displaying almost the same trend up to a vertical
shift. The highest proportion of IPDs exceeding the ''maximal'' value
was present in the nocturnal nature scene. This was most probably
caused by a large number of very similar sound sources
(grasshoppers) at each side of the 
head. They generated non-synchronized and strongly overlapping
waveforms. Phase information in each ear resulted therefore from an
acoustic summation of multiple sources, hence instantenous IPD was not
directly related to a single source position and often exceeded the
$IPD_{max}$ value. Surprisingly, IPDs in the most spatially dynamic
scene - city center - did not exceed the $IPD_{max}$ limit as
often. This may be due to a smaller number of sound sources present
and may indicate that the proportion of ''forbidden'' IPDs is a
signature of a number of sound sources present in the scene. For nocturnal nature and city center scenes the proportion peaked at $400$ Hz achieving values of $0.45$ and $0.35$ respectively. For a forrest walk scene, the peak at $400$ Hz did not exceed the value of $0.31$ at $200$ Hz. All proportion curves converged to $0$ at $734$ Hz frequency, where $IPD_{max} = \pi$.

\subsubsection*{Separation of speech with single channel IPDs}
\label{vmMixtures}
As already mentioned before, IPD distributions at most  frequency
channels in the forrest walk scene revealed an additional property,
namely a clear, sharp peak at $0$ radians. This feature was not
present in the two other, statically recorded scenes. As an example,
IPD distribution at $561$ Hz is depicted in figure \ref{fig:IPD4}
A. The histogram structure reflects the elevated presence of sounds with IPDs close to $0$ hence equal monaural phase values. Zero IPDs can be generated either by sources located at the midline (directly in front or directly in the back) or self-produced sounds such as speech, breathing or loud footsteps.

As visible in figure \ref{fig:IPD4} two components contributed to the structure of the marginal IPD distribution - the sharp ''peak component'' (dashed blue line) and the broad ''background'' (dashed red line). Due to this property, IPD histograms were well suited to be modelled by a mixture model. This means that their pdf could be represented as a linear combination of two von Mises distributions in the following way
\begin{equation}
\label{vmMixture}
p(IPD_\omega | \mathbf{\kappa_\omega}, \mathbf{\mu_\omega}) = \sum_{i=1}^2 p(C_i) p(IPD_\omega | \kappa_{\omega,i}, \mu_{\omega,i})
\end{equation}
where $\mathbf{\kappa_\omega} \in \mathbb{R}^2$ and
$\mathbf{\mu_\omega} \in \mathbb{R}^2$ are parameter vectors, $C_i \in
\{1, 2\}$ are class labels, $p(C_i)$ are prior probabilities of class
membership and $p(IPD_\omega | \kappa_{\omega,i}, \mu_{\omega,i})$ are
von Mises distributions defined by equation \ref{eq:IPD2}. A fitted
mixture of von Mises distributions is also visible in figure
\ref{fig:IPD4} A, where dashed lines are mixture components and a
continuous black line is the marginal distribution. It is clearly
visible that a two-component mixture fits the data much better than a
plain von Mises distribution. There is also an additional advantage of
fitting such a mixture model, namely it allows to perform a classification problem and assign each IPD sample (and therefore each associated sound sample) to one of the two classes defined by mixture components. Since the prior over class labels is assumed to be uniform, this procedure is equivalent to finding a maximum-likelihood estimate $\hat{C}$ of $C$
\begin{equation}
\label{eq:decoding}
\hat{C} = \arg \max_C p(IPD_\omega | C)
\end{equation}
In this way, if no sound source at the midline is present, a separation of self generated sounds from the background should be easily performed using information from a single frequency channel. Results of a self-generated speech separation task are displayed in figure \ref{fig:IPD4} B. A two-second binaural sound chunk included two self-spoken words with a background consisting of a flowing stream. Each sample was classified basing on an associated IPD value at $561$ Hz. Samples belonging to the second, sharp component are coloured blue and background ones are red. It can be observed that the algorithm has successfully separated spoken words from the environmental noise. Audio samples are available in the supplementary material. 

\subsection*{Independent components of binaural waveforms}

In this section, instead of studying predetermined features of the stimulus (binaural cues), we use binaural waveforms to train Independent Compnent Analysis (ICA) - a statistical model which optimizes a general-purpose objective - coding efficiency \cite{BellSejnowski}. In the ICA model, short ($8.7$ ms) epochs of binaural sounds are assumed to be a linear superposition of basis functions multiplied by linear coefficients $s$ (see figure \ref{fig:ICA1} A). Linear coefficients are assumed to be independent and \emph{sparse}, i.e., close to $0$ for most of data samples in the training dataset. Basis functions learned by ICA can be interpreted as patterns of correlated variability present in the dataset. 

Figure \ref{fig:ICA1} B depicts exemplary basis functions learned from
each recording. Each feature consists of two parts, representing
signal in the left and the right ear (black and red colours
respectively). Features trained on different recordings vary in their
shape. Those differences are explicitely visible in spectrotemporal
representations of basis functions depicted on figure
\ref{fig:ICA2}. Each shape corresponds to an equiprobability contour
of a Wigner distribution associated with a single basis
function. Wigner distributions localize energy of a temporal signal in
the time frequency plane. Left and right ear parts belonging to the
same feature are plotted with the same color. The obtained time-frequency tilings reveal a strong dependence on the auditory scene. Firstly basis function shapes are different - from time extended and frequency-localized in the city center scene, to temporally brief, instantenous features of the forrest walk scene. Despite shape differences, in each case, basis functions tile the time-frequency plane uniformly. Their shapes constitute an interesting aspect of the auditory scene and can be compared with results obtained by \cite{AbdallahPlumbey, Lewicki}. This is, however not the focus of the current work.

Sounds of the most spatially static scene - nocturnal nature - were modelled mostly by features of the same spectrotemporal properties in each ear (with an anomally which occurred around $3.5$ kHz). This is visible in figure \ref{fig:ICA2} - blobs of the same color lie mostly in the same region on the left and the right ear plots. 
In more dynamic scenes, independent components (ICs) captured different, non-trivial dependencies. Pure frequency features learned from the city center recording had similar monaural parts below $3.5$ kHz. Above this threshold, a cross-frequency
interaural coupling appeared - in the right ear panel, blue colored
features lie in the high frequency regime, while in the left ear  they
occupy a low frequency region. This means that to represent natural binaural signal efficiently, monaural information from different frequencies should be processed simultaneously. 
Interaural dependencies represented by ICs of the forrest walk scene were even more complex. Since most of the basis functions were much more temporal than spectral, time dependencies were also captured in addition to the spectral ones. High frequency events in the right ear were coupled with more temporally extended, low-frequency features of the left ear. Interestingly, tiling of the time-frequency plane associated with the right ear was not as uniform as for the left one. 

The majority of learned basis functions was highly localized in
frequency, which agrees with results obtained by
\cite{AbdallahPlumbey, Lewicki, SmithLewicki}. However, some basis
functions did not have well localized spectra. They were excluded from
the analysis, that is why the number of basis functions varies across
the analyzed auditory scenes. See materials and methods for the detailed discussion.
To understand how spectral power was distributed in monaural parts of ICs, we computed a peak power ratio (PPR):
\begin{equation}
\label{peakPower}
PPR = 10\log_{10}(\frac{A_{max, L}}{A_{max, R}})
\end{equation}
where $A_{max, L}, A_{max, R}$ are maximal spectrum values of the
left and right ear parts of each IC respectively. Each circle in
figure \ref{fig:ICA3} represents a single IC. Its vertical and
horizontal coordinates are monaural peak frequencies and colors encode
the PPR value. Features which lie along the diagonal can be considered
as a representation of ''classical'' ILDs, since they encode features
of the same frequency in each ear and differ  only in level. ICs lying
away from the diagonal with high absolute PPR values represent more
monaural information, and those with the low absoulte PPR other
aspects of the stimulus, such as interaural cross-frequency
couplings. Figure \ref{fig:ICA4} depicts proportion of features of
same monaural frequencies (on diagonal) and those which bind different
frequency channels (off diagonal). A pronounced difference among
auditory scenes is visible in figure \ref{fig:ICA3}. The majority of basis functions learned from the nocturnal nature scene ($161$) clusters closely to the diagonal. 
The basis function set trained on the mostly dynamic scene (city center) separates into three clear subpopulations. Two of them, including $140$ features were monaural. Monaural basis functions were dominated mostly by the spectrum of a single ear part, and the part representing the contralateral ear was of a very low frequency, close to a DC component. The binaural subpopulation contained $111$ basis functions perfectly aligned with the diagonal. Such separation suggests that waveforms in both ears were highly independent and should be modelled using a large set of separate, monaural events.
ICA trained on the forrest walk scene yielded a set of basis functions
which was a compromise between nocturnal nature and city center
scenes. Even though the highest number of features - $165$ lied off
the diagonal, the separation was not as sharp as for the city center
scene. What clearly appeared was a division into two subpopulations,
members of which were dominated by the spectrum of one of the
ears. ICs mostly coupled low frequencies ($<2$ kHz) from one ear with
a broad range of frequencies in the other. Those properties may imply
that in the case of this scene, both - features modelling binaural
dependendencies and capturing purely monaural events  -  were required to model the data. 
To allow further comparison of learned ICs and known coding mechanisms
in the binaural auditory system, we computed ILD and IPD cue
values. This was done only  for features encoding the same
frequency information in both ears, since phase differences are ill
defined otherwise, and auditory brainstem extracts cues mostly from
the same frequency channels \cite{GrotheMcAlpine}. Results are visible
in figure \ref{fig:ICA5}. IPDs represented by independent components
separated into two channels in the city center and forrest walk
scenes. The range of IPDs was higher for a more spatially varying
scene, which is visible as a strong scatter of points. 

For the nocturnal nature scene no such separation is visible. This is perhaps
due to the fact that object positions were mostly fixed, generating
lowly-varying IPDs captured by the learned ICs. Therefore the model
did not have to generalize over a broader range of IPDs. ILDs in turn,
in all scenes were separated into two distinct channels. The
separation strength correlated with the scene's spatial variability and was 
highest for the city center scene and lowest for the nocturnal nature. Interestingly, in the latter one, ILD features were present also in high frequencies which was not the case in the two others. Also here, the separation of features seems to reflect the spatial structure and dynamics of the auditory scene.

\section*{Discussion}

Binaural cues are usually studied in a relationship to the angular position of the generating stimulus \cite{FischerNature, Fischer, Hofman}. In probabilistic terms this corresponds to modelling the \emph{conditional} probability distribution $p(cue|\theta)$, where $\theta$ is the angular stimulus location. According to the Bayes theorem, position inference given the cue can be then performed by: (a) computing the posterior distribution $p(\theta|cue)$ and (b) identifying $\theta$ for instance, as a maximum of the posterior distribution. Formally this process can be described by the following equations:

\begin{equation}
\label{cuePosterior}
p(\theta|cue) \propto p(cue|\theta) p(cue)
\end{equation}

\begin{equation}
\label{cueMap}
\hat{\theta} = \arg \max_\theta p(\theta|cue)
\end{equation}
where $\hat{\theta}$ is the estimated position. The Bayesian approach to sound localization has been succesfuly applied before, for instance to predict behavior and neural representation of binaural cues in the barn owl \cite{FischerNature}.

In the present study, we focused on \emph{marginal} distributions of
cues and binaural waveforms. This approach allows us to understand aspects of binaural hearing in the natural environment which are not directly related to the sound localization task. Marginal distributions $p(cue)$  describe global properties of the stimulus to which the nervous system is exposed under natural conditions.  Knowledge of a typical stimulus structure allows to predict properties of the sensory neurons \cite{OlshausenSimoncelli, BarlowUnsup} and helps in understanding the complexity of the task such as binaural auditory scene analysis, when performed in ecological conditions.

\subsection*{Binaural cues in complex auditory environments}

Binaural scenes recorded and studied in this paper were selected to
represent broad groups of possible auditory environments of different
acoustic and spatial properties. In all three cases, waveforms in each
ear were for most of the time an acoustic summation of multiple sound
sources. Additional factors, which influenced monaural stimuli were
motion trajectories of objects and the listener, as well as sound
reflections. Instantenous binaural cue values were therefore not
generated by a single, point source, but were a function of a complex
auditory scene. Inversion of a cue value to a sound position becomes,
in such a setting, an inverse problem, since multiple scene
configurations could give rise to the same cue value (for instance an
ILD equal to $0$ can be generated by a single source located at the
midline, or two identical sources symmetricaly located on both sides
of the head, see section \ref{sec:ILD}). In such scenarios, the sound
localization task can not be performed as a simple inversion of a cue
value to the sound position (the most simple case described by
equations \ref{cuePosterior} and \ref{cueMap}). It rather becomes
equivalent to the \emph{cocktail party problem}
\cite{McDermottCocktail}. Localization of a sound source in complex
listening situations has been a subject of substantial psychophysical
\cite{Blauert} and electrophysiological \cite{Takahashi,
  KellerTakahashi, Day, Baxter} research. An interesting theoretical
model has been suggested by Faller and Merimaa
\cite{FallerMerimaa}. The authors of this study proposed that to
localize one sound source out of many present, the auditory system
could use instantenous binaural cues only in time intervals when the left
and the right ear waveforms are highly coherent (i.e. their
cross-correlation peak exceeds a certain threshold). In such brief
moments, ILD and IPD values would correspond to only a single
source. This mechanism is able to explain numerous psychophysical
studies. Meffin and Grothe \cite{MeffinGrothe}  hypothesized that the
auditory brainstem may perform low-pass filtering of localization cues
to reject rapidly fluctuating ''spurious'' cue values, which may
originate from multiple sources. The aforemetioned mechanisms, however, involve a rejection of a large amount of information, by discarding ''ambiguous cues'', which may  still contain information useful in the auditory scene parsing. 
In very general terms, a useful strategy for the auditory system would be to use higher dimensional stimulus features (such as temporal cue sequences, or cross-frequency cue dependencies) to separate a source (or sources) of interest from the background and infere its spatial configuration. It has been demonstrated, for instance, that neurons in the Inferior Colliculus of the rat show a stronger response to dynamic, ''ecologically valid'' IPD sequences, than to constant IPDs \cite{SpitzerSemple1, SpitzerSemple2}. In the auditory cortex of macaque monkeys, neurons become sensitive to even more complex IPD sequences  \cite{Malone}. Such properties may be examples of tuning to high-dimensional, binaural stimulus aspects. 
In the above mentioned view, instantenous binaural cues, as extracted by the early brainstem nuclei LSO and MSO \cite{GrotheMcAlpine}, provide information useful in the auditory scene analysis task. In natural conditions however, their mere identification is not necessarily equivalent to the localization of the sound position. Binaural cues may rather serve as inputs to further computations (which are not necesserily limited to sound localization per se) performed in the higher stages of the binaural auditory pathway.

\subsection*{Implications for neural processing and representation of binaural sounds}

As predicted by physics of sound propagation, monaural phase values in
natural environments reveal dependence in their difference. The strength of the dependence is measured by $\kappa$ - the concentration parameter of the von Mises IPD distribution \cite{CadieuPhase}. Interestingly, humans stop using IPDs to localize sounds above $1.5$ kHz \cite{WightmanKistler} i.e. the frequency regime, where monaural phases become marginally independent (as reflected by the decay of the $\kappa$ parameter). 

In anechoic environments, point sources of sound generate ITD values
which are constrained by the head size of the listener. It has been,
however, observed that in many species, IPD sensitive neurons have
peaks of their tuning curves located outside of this ''physiological''
range \cite{GrotheMcAlpine}. This representational strategy has been
explained by suggesting that in mammals IPDs are encoded by the activity
of two separate, broadly tuned neural channels. Notably, such a representation emerges as a consequence of maximizing Fisher information about naturally occuring IPDs \cite{HarperMcAlpine}. Here, we demonstrate that in natural hearing conditions a substantial amount of IPDs (up to $45 \%$) lies outside of the physiological range. Those IPD values may be a result of a reflection \cite{GourevichBrette} or a presence of multiple spatially separate desynchronized sound sources \cite{GrotheMcAlpine}. Sound reflections generate reproducible cues and carry information about the spatial properties of the scene \cite{GourevichBrette}. If a large IPD did not arise as a result of a reflection, it means that at least two sound sources contribute to the stimulus at the same frequency. Especially in the latter case, IPDs provide not only spatial information useful to identify the position of the sound, but become a strong source separation cue.
Proportion of IPDs exceeding the physiological range decreased with growing frequency (since the maximal IPD limit increases). This observation agrees with the experimental data showing that in many species, neurons with low best frequency are tuned to large IPDs which often exceed the physiological range \cite{McAlpineJiangPalmer, BrandBehrend, Hancock, Kuwada}. Taken together, IPDs larger than predicted by the head size occur frequently in natural hearing conditions and carry important information. This can be an additional factor, explaining why in mammals, peaks of the IPD tuning curves lie outside of the physiological range.
As demonstrated in section \ref{vmMixtures}, interaural phase differences can be used for not explicitely spatial hearing tasks, such as extraction of self-generated speech (and potentially other sounds, such as steps). If there is no sound source present at the midline location (corresponding to $0$ IPD), a simple classification procedure suffices to identify and separate vocalization of oneself from the background sound using information from a single frequency channel. Differentiation between self generated sounds and sounds of the environment is a behaviorally relevant task which has to be routinely performed by animals.
   
According to the Duplex Theory, ILDs contribute mostly to localization of high frequency sounds, since the head attenuates higher frequencies much stronger than lower ones \cite{Blauert}. Analysis of human Head Related Transfer Functions (HRTFs) shows that ILDs are almost constant at different spatial positions for low frequencies and become more variable (hence more informative) when frequency increases above $4$ kHz \cite{TheShapeOfEars}. For single sound sources in anechoic environments, ILDs can have values as large as $40$ dB \cite{TheShapeOfEars}.
Based on those observations, one could expect that natural ILD distributions
are strongly frequency dependent. Somewhat surprisingly, natural
distributions reveal a quite homogenous structure across different frequency channels, which is well captured by the logistic distribution. Overall, averages are equal (or very close to) $0$ dB in different auditory environments, and for all studied frequencies. Variance slightly increases with increasing frequency. Homogeneity of distribution forms and averages can be explained in the following way. Typical, natural auditory scenes consist of similar sound sources at both sides of the head (human speakers, grasshoppers, wind, etc). Each of the sound sources has a similar spectrum, hence they all contribute to waveforms in the left and right ear mutually cancelling each other. For this reason, ILD averages are close to $0$ and have a similar shape in different environments. The variance increase can be explained by properties of the head related filtering since small movements of high frequency sources give rise to a large ILD variability. 
As mentioned before, interaural level differences are mostly believed to contribute to localization of high frequency sounds \cite{Blauert} since then they are large enough to be easily detectable. It has been, however, demonstrated that sound sources proximal to the listener can generate pronounced ILDs also in low frequencies (below $1.5$ kHz) \cite{BrungartRabinowitz, ShinnCunningham}. Our results show that in natural environments the auditory system is exposed to a similar ILD distribution across all frequencies, including the low ones. The distribution includes also relatively large values (above $10$ dB).  Close sound sources and other environmental factors such as the wind perceived in only one ear generate large low-frequency ILDs. One could therefore speculate that neurons with low best frequencies should also form an ILD representation. Indeed, such neurons have been found in the Lateral Superior Olive (LSO) of the cat \cite{TollinYin}.

To go beyond studying one-dimensional features of the binaural signal
(ILDs and IPDs), the probability distribution of short binaural
waveforms was modelled by performing Independent Component Analysis. A
similar analysis in the visual domain was performed by Hoyer and
Hyv\"{a}rinen \cite{HoyerStereo} for binocular image pairs. The ICA
algorithm has identified complex patterns of dependency, different for
each studied auditory scene. Interestingly, the spectrotemporal shape
of the  monaural parts of the  basis functions varied strongly across
recorded auditory scenes. The obtained results can be compared with other studies which applied Independent Component Analysis to natural sounds \cite{Lewicki, AbdallahPlumbey, BellSejnowskiSound}. 
Linear codes learned by the ICA model show that one should adopt
different representations depending on the properties of the acoustic
environment. A static scene (nocturnal nature) generated monaural
waveforms, which were highly redundant, since the signal in each ear
was originating mostly from the same source. For this reason, the majority of basis functions represented amplitude fluctuations in both ears and in the same frequency channels. When sound sources moved rapidly and independently from each other at both sides of the head, waveforms in each ear were much less redundant. That is why a representation of the dynamic binaural scene (city center) consisted of three, clearly separate populations of basis functions - two representing monaural signal, and one binaural. Interestingly, binaural functions coupled monaural channels of the same frequency. The moderatly dynamic scene (forrest walk) was best represented by basis functions which were mostly monaural and modelled a broad range of binaural cross-frequency dependencies.  
A variety of different dependency forms were captured, including temporal, spectral and spectrotemporal ones. This implies that information present in the binaural signal goes beyond instantenous binaural cue values. This notion goes in line with studies, which have found and characterized spectrotemporal binaural neurons at the higher stages of the auditory pathway \cite{Qiu, Miller}. Binaural hearing in the natural environment may also rely on comparison of spectrotemporal information at both sides of the head. 

\subsection*{Conclusions}

In the present study, we analyzed marginal statistics of binaural cues and waveforms. Thereby, we provided a general statistical characterization of the stimulus processed by the binaural auditory system in natural listening conditions. We have also made availible natural binaural recordings, which may be used by other researchers in the field.
In a broad perspective, this study contributes to the lines of research that attempt to explain properties of the auditory system by analyzing natural stimulus structures. Further understanding of binaural hearing mechanisms will require a more systematic analysis of higher order stimulus statistics. This is the subject of future research.

\section*{Materials and Methods}
\subsection*{Recorded scenes}

The main goal of the study was to analyze cue distributions in different auditory environments. To this end, three auditory scenes of different spatial dynamics and acoustic properties were recorded. Each of the recordings lasted $12$ minutes.
\begin{enumerate}
\item \textbf{Nocturnal nature} - the recording subject sat in a
  randomly selected position in the garden during a summer evening. During the recording the subject kept his head still, looking ahead, with his chin parallel to the ground. The dominating background sound are grasshopper calls. Other acoustic events included sounds of a distant storm and a few cars passing by on a near-by road. The spatial configuration of this scene did not change much in time - it was almost static. 

\item \textbf{City center} - the recording subject sat in a touristic area of an old part of town, fixating the head as in the previous case. During the recording many moving and static human speakers were present. Contrasted with the previous example, the spatial configuration of the scene varied continuously.

\item \textbf{Forrest walk} - this recording was performed by a
  subject freely moving in the wooded area. A second speaker was
  present, engaged in a free conversation with the recording
  subject. In addition to speech, this scene included environmental
  sounds such as flowing water, cracks of broken sticks, leave
  crunching, wind etc. The binaural signal was affected not only by the spatial scene configuration, but also by the head and body motion patterns of the recording subject.
\end{enumerate}

Two of the analyzed auditory scenes (nocturnal nature and city center)
were recorded by a non-moving subject, therefore sound statistics were
unaffected by the listener's motion patterns and self generated sounds. In the third scene (forrest walk) the subject was moving freely and speaking sparsely. Scene recordings are available in the supplementary material.

\subsection*{Binaural recordings}

Recordings were performed using  Soundman OKM-II binaural microphones
which were placed in the left and the right ear channels of the
recording subject.  A Soundman DR2 recorder was used to simultaneously record sound in both channels in an uncompressed wave format at $44100$ Hz sampling rate. The head circumference of the recording subject was equal to $60$ cm. Assuming a spherical head model this corresponds to a $9.5$ cm head radius.

\subsection*{Frequency filtering and cue extraction}

\begin{figure}[h]
\includegraphics{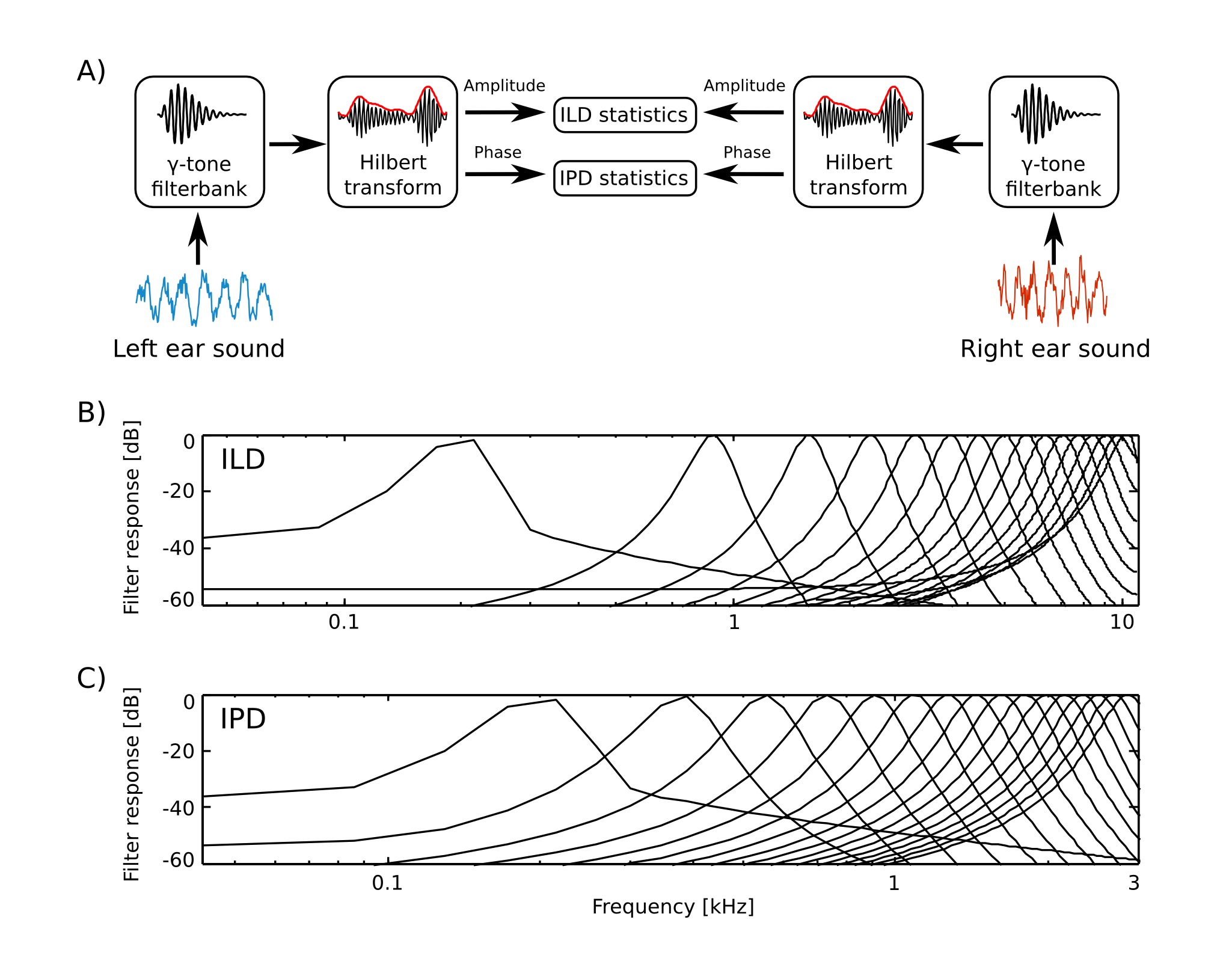}
\caption{Preprocessing and cue extraction pipeline}
\label{schema}
\end{figure}

\begin{figure}[h]
\includegraphics{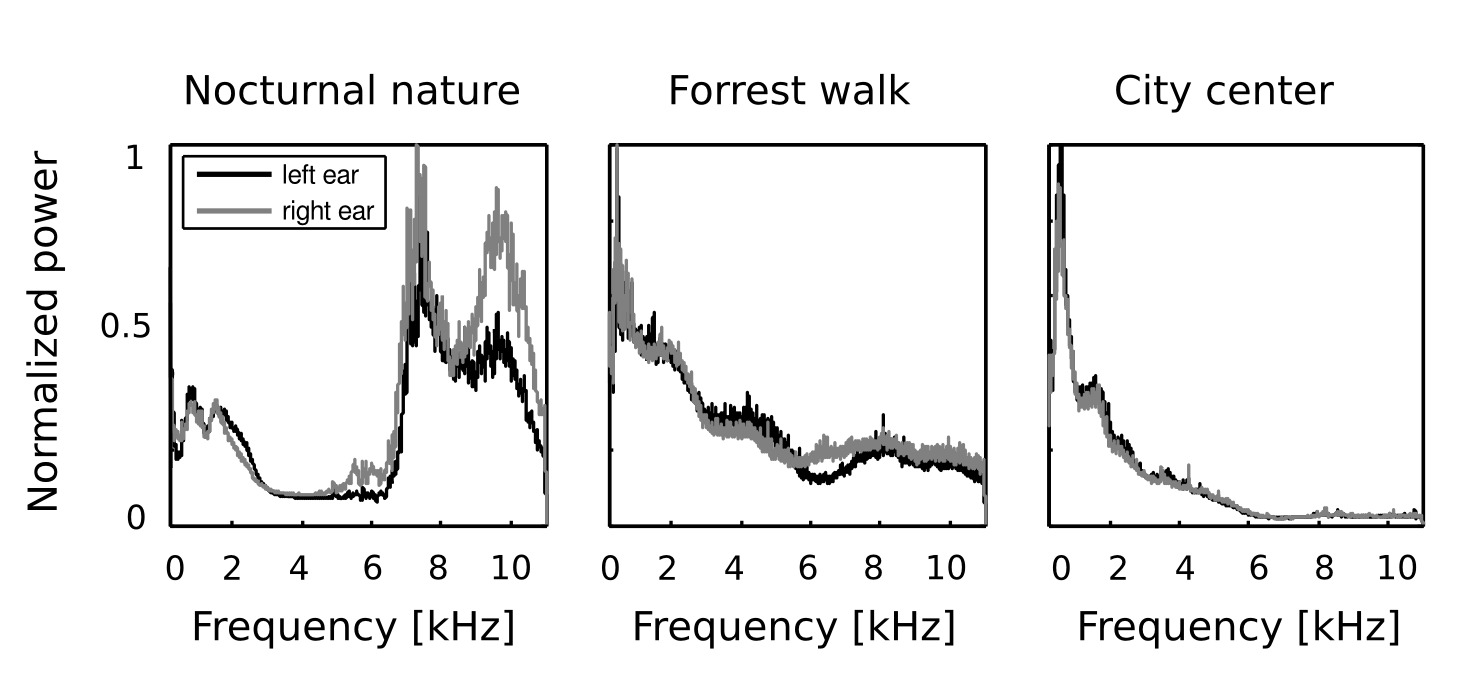}
\caption{Frequency spectra of binaural recordings}
\label{fig:spectra}
\end{figure}

\begin{figure}
\centering
\includegraphics{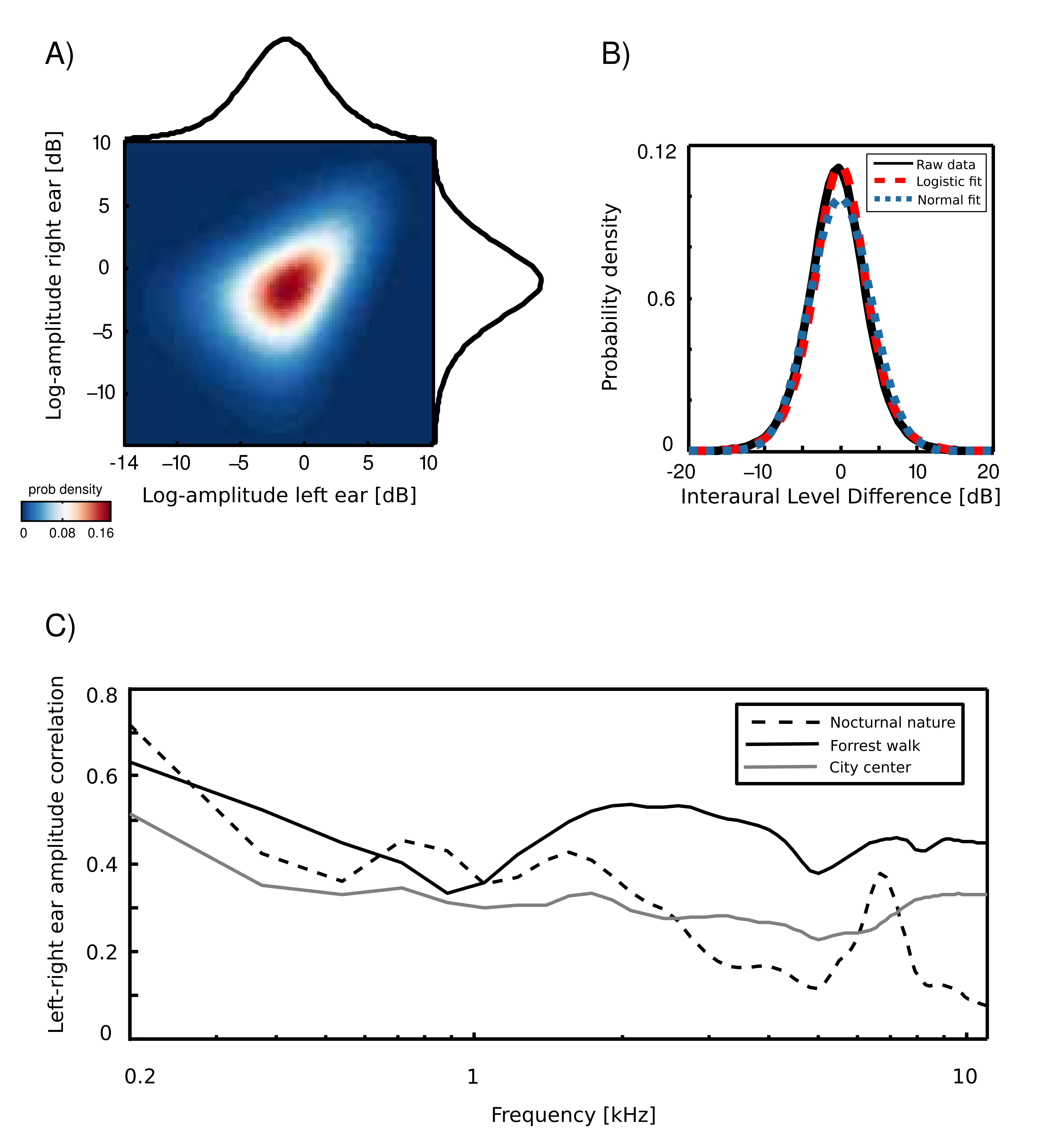}
\caption{Binaural amplitude statistics. A) An exemplary plot of joint amplitude distribution in both ears B) ILD distribution for a fixed channel toghether with a Gaussian and a logistic fit C) Interaural correlations of amplitudes across frequency channels}
\label{fig:ILD1}
\end{figure}

\begin{figure}
\centering
\includegraphics{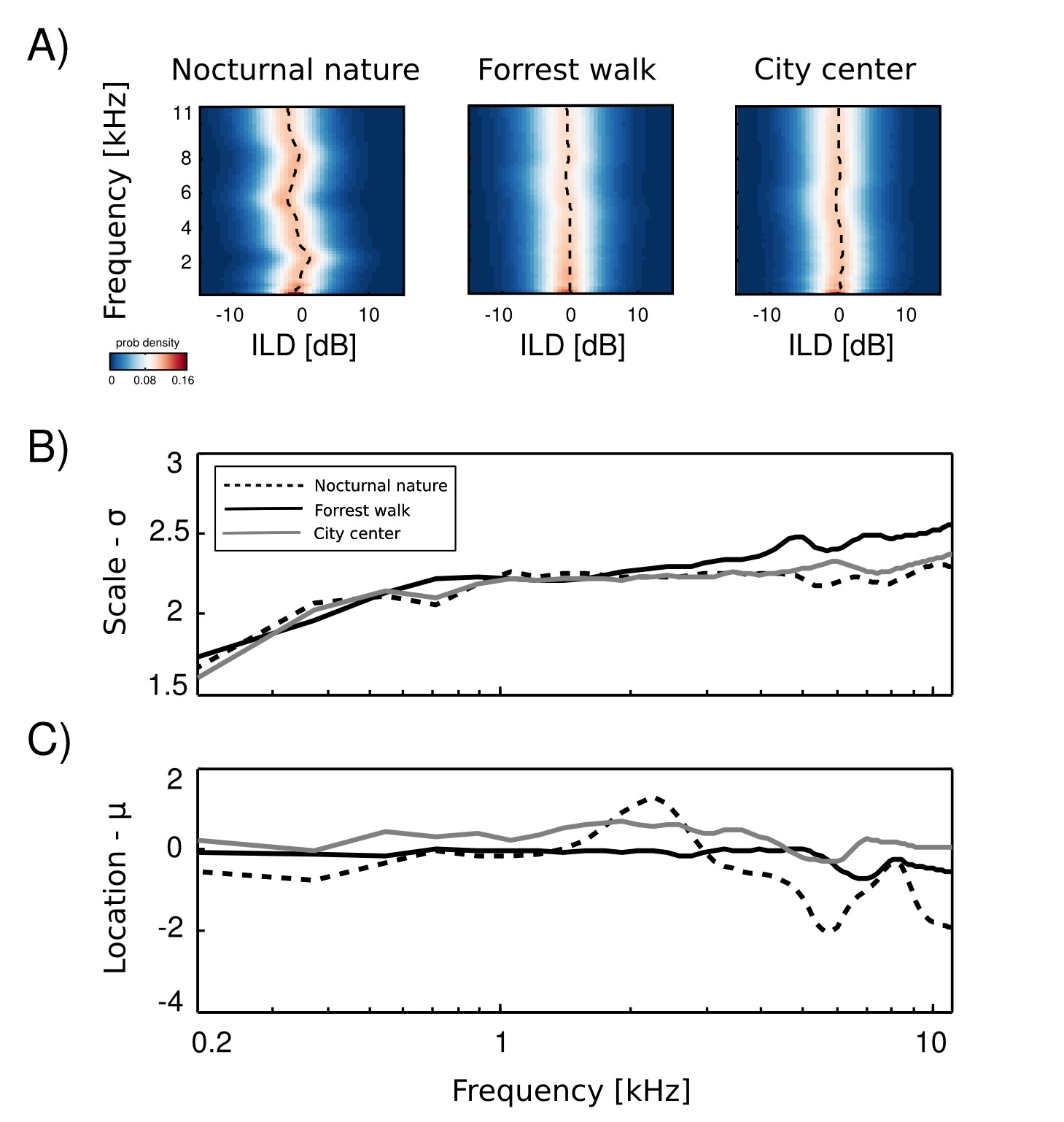}
\caption{Interaural level difference distributions. A) Histograms plotted as a function of frequency B) }
\label{fig:ILD2}
\end{figure}

\begin{figure}
\centering
\includegraphics{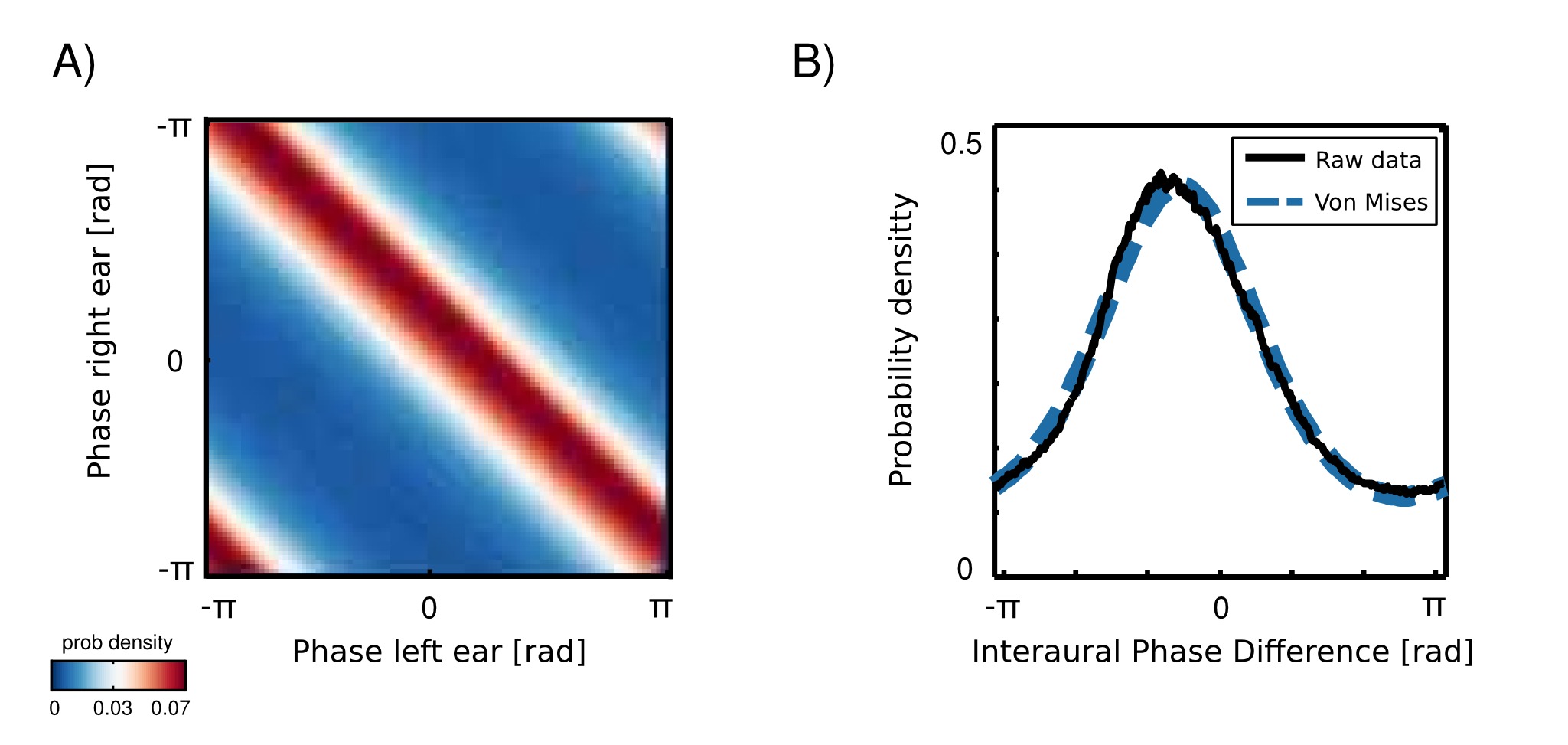}
\caption{Binaural phase statistics A) Exemplary joint probability distribution of monaural phases B) An IPD histogram (black line) and a fitted von-Mises distribution}
\label{fig:IPD1}
\end{figure}

\begin{figure}
\centering
\includegraphics{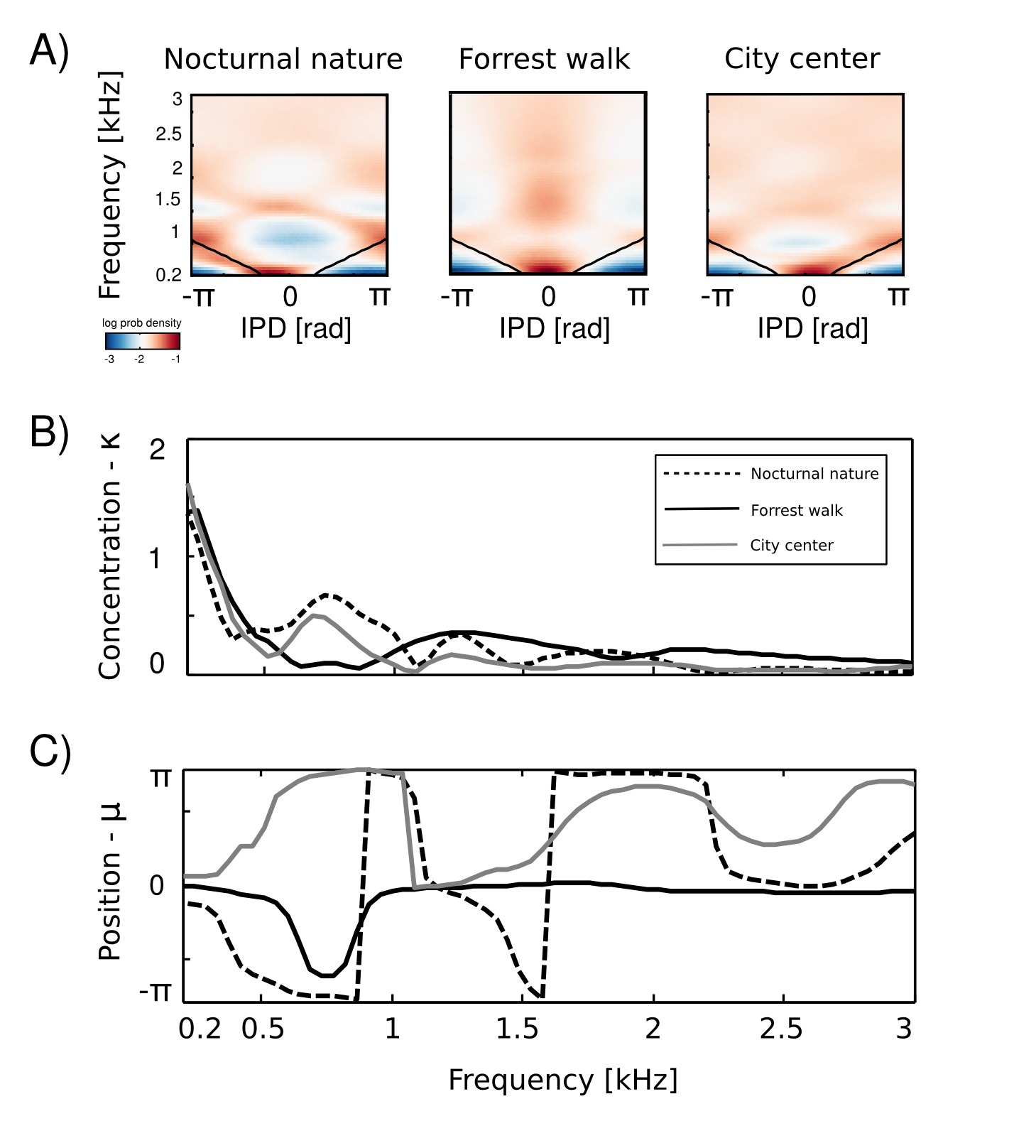}
\caption{IPD distributions. A) Histograms B) Concentration parameter $\kappa_\omega$ as a function of frequency C) Position parameter $\mu_\omega$ as a function of frequency}
\label{fig:IPD2}
\end{figure}

\begin{figure}
\centering
\includegraphics{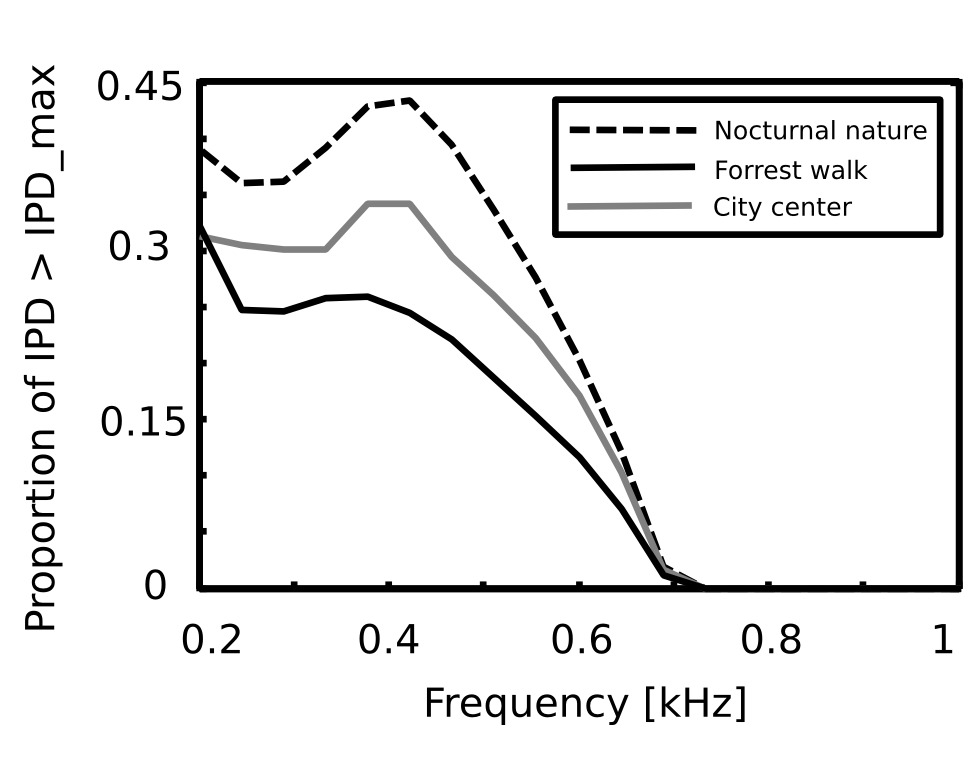}
\caption{Proportion of IPDs exceeding the ''maximal IPD'' threshold in each frequency channel}
\label{fig:IPD3}
\end{figure}

\begin{figure}
\centering
\includegraphics{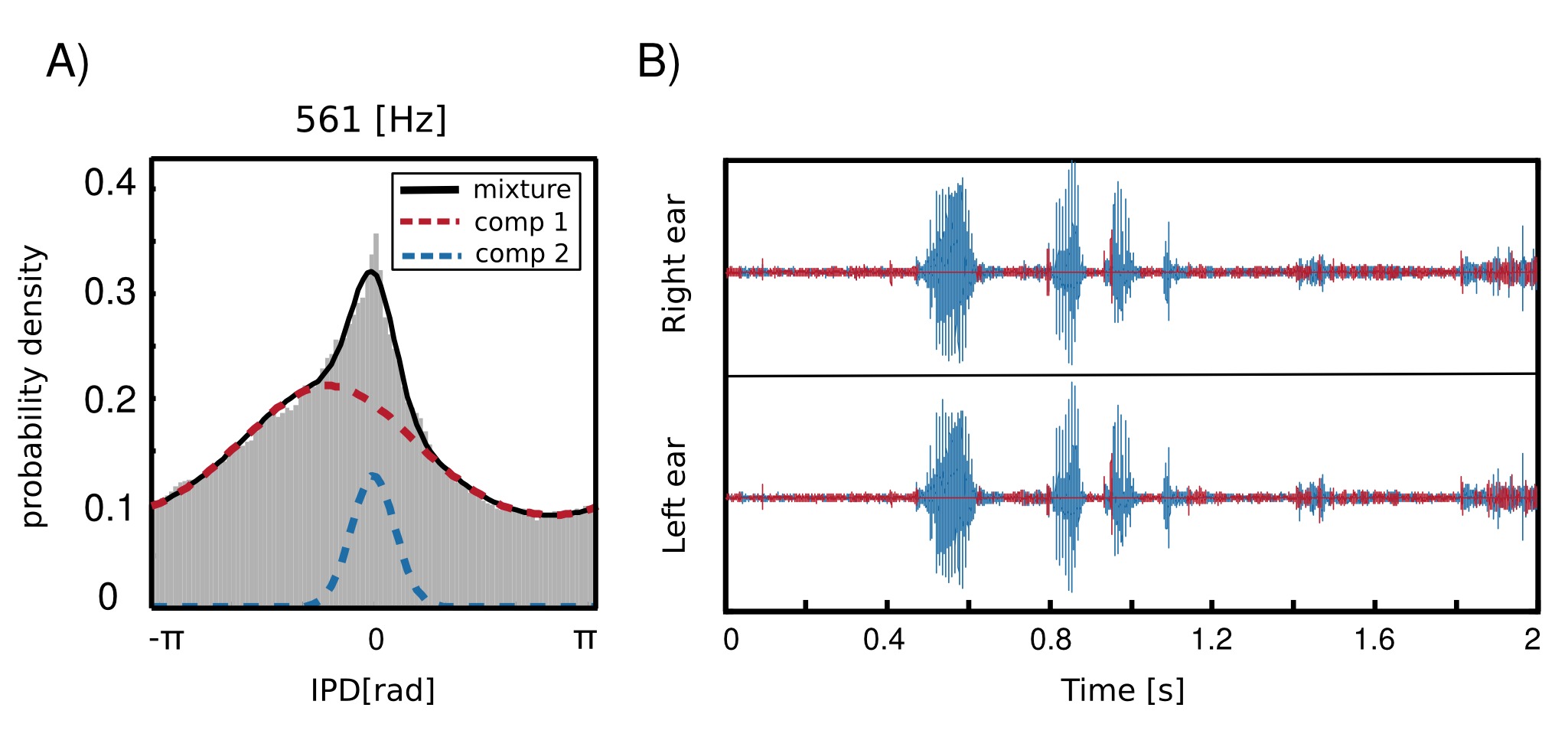}
\caption{Self speech separation using single channel IPDs. A) An exemplary IPD distribution in the forrest walk scene B) Classification results}
\label{fig:IPD4}
\end{figure}

\begin{figure}
\centering
\includegraphics{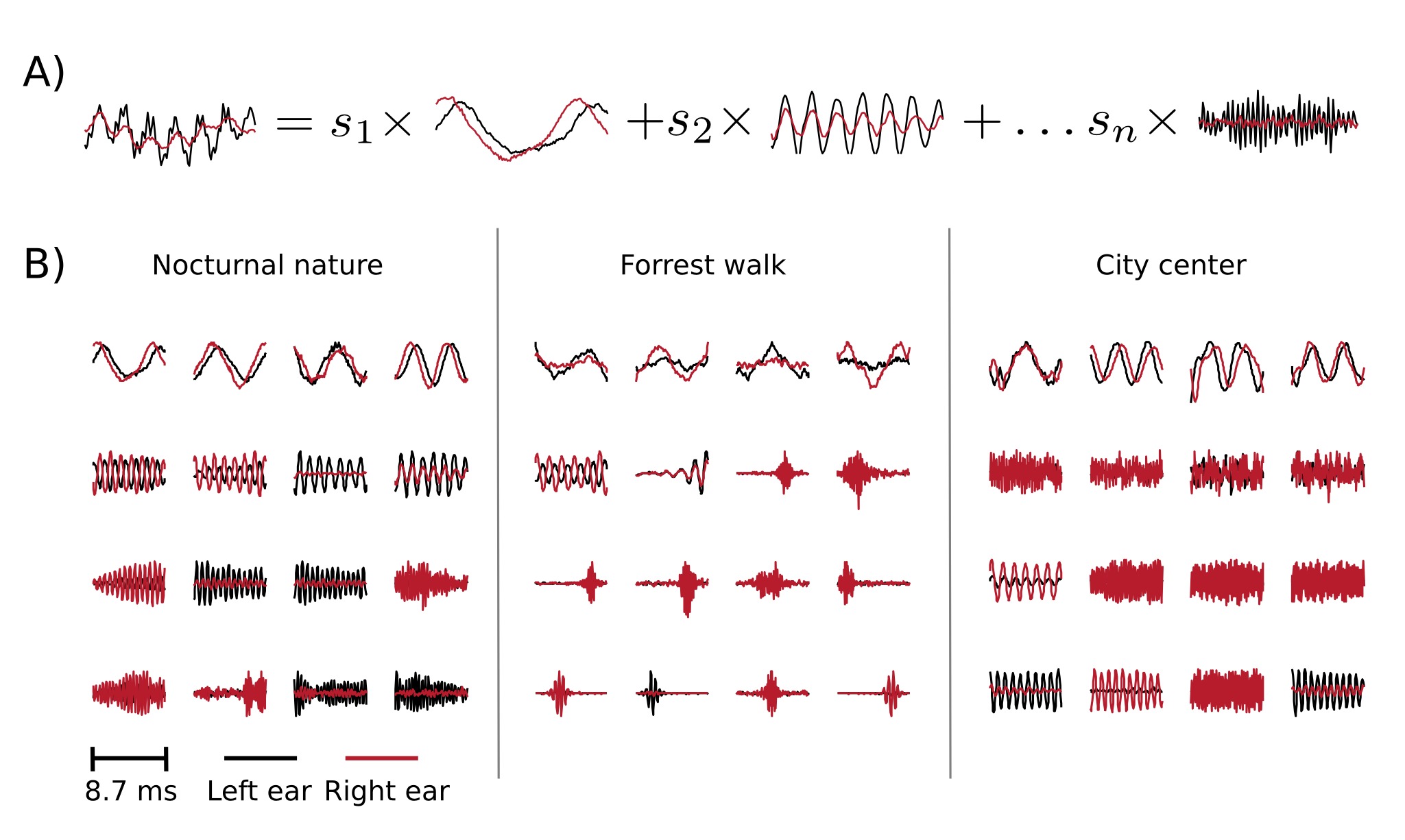}
\caption{Independent components of natural binaural sounds. A) Explanation of the ICA model. Coefficients $s_i$ are assumed to be sparse and independent. B) Exemplary ICA basis functions from each recorded scene.}
\label{fig:ICA1}
\end{figure}

\begin{figure}
\centering
\includegraphics{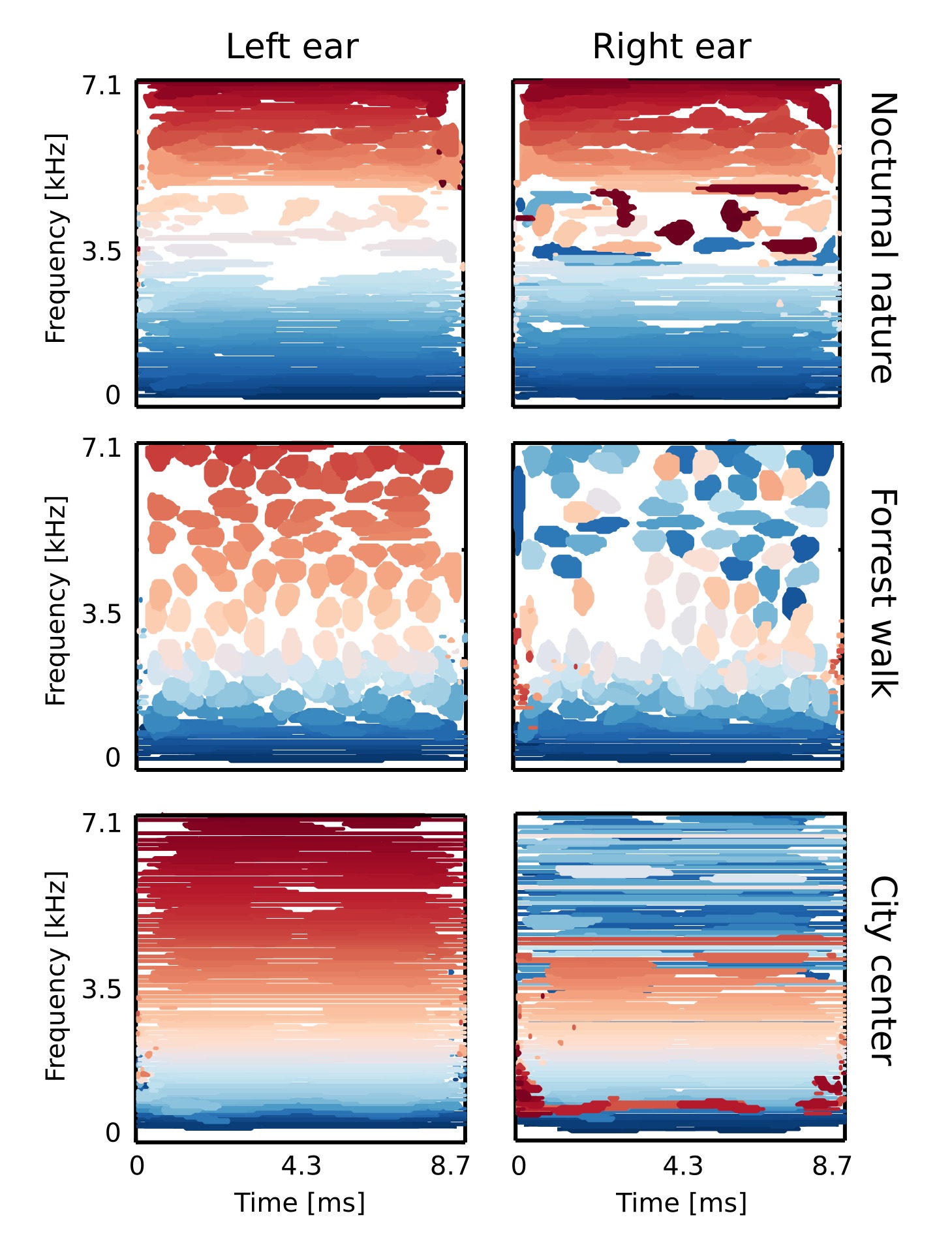}
\caption{Independent components plotted on a time frequency plane. Rows correspond to auditory scenes. Columns to ears. Shapes of the same color form a single independent component.}
\label{fig:ICA2}
\end{figure}

\begin{figure}
\centering
\includegraphics{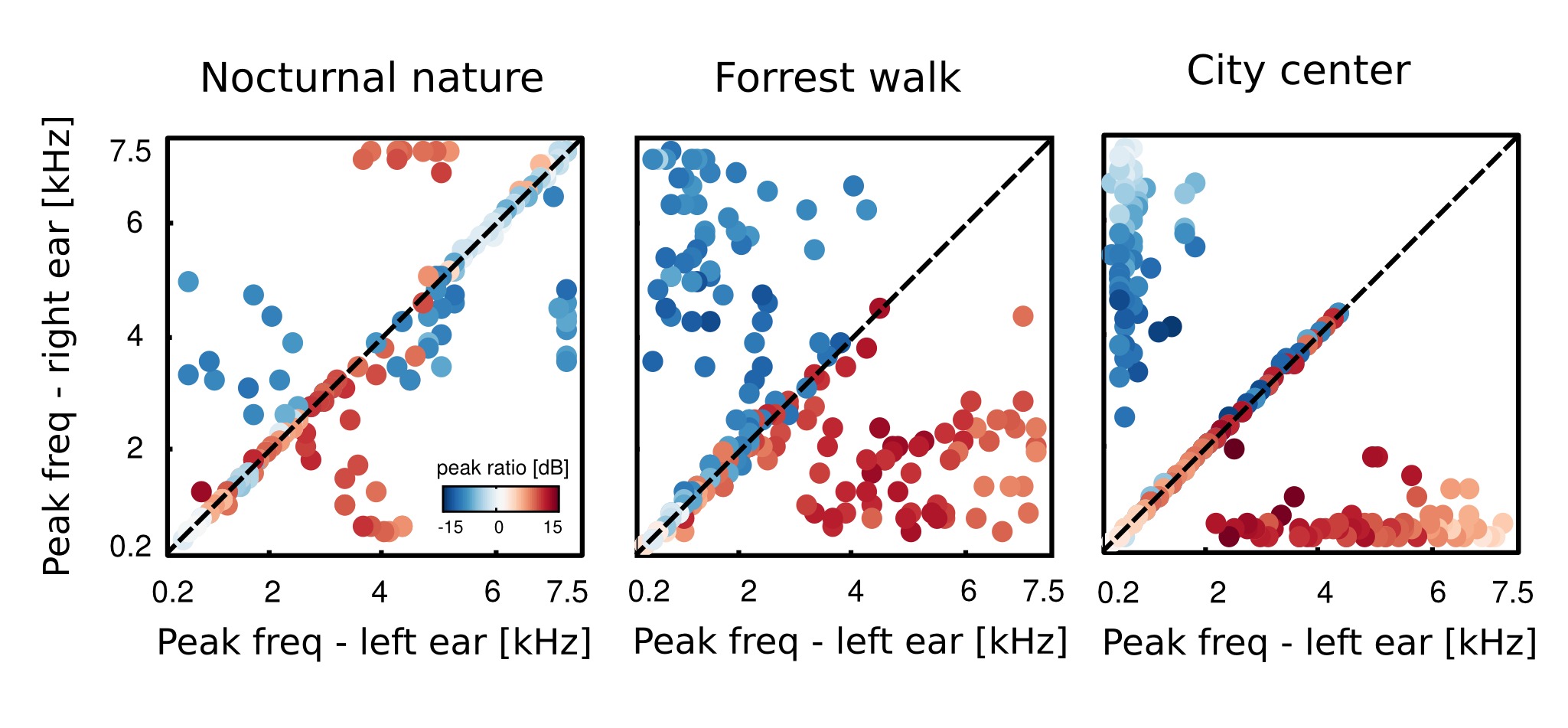}
\caption{Peak frequencies of IC monaural parts plotted against each other. Colors encode the Peak Power Ratio}
\label{fig:ICA3}
\end{figure}

\begin{figure}
\centering
\includegraphics{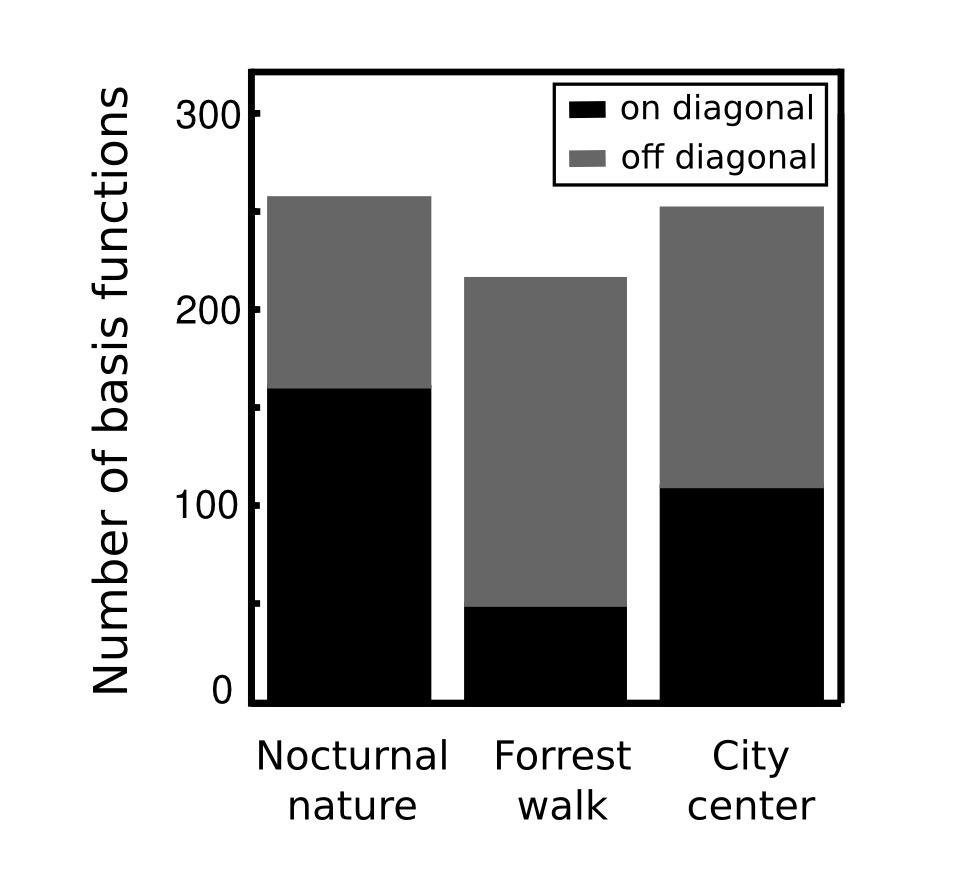}
\caption{Proportion of Independent Components with the same frequency peak in each ear}
\label{fig:ICA4}
\end{figure}

\begin{figure}
\centering
\includegraphics{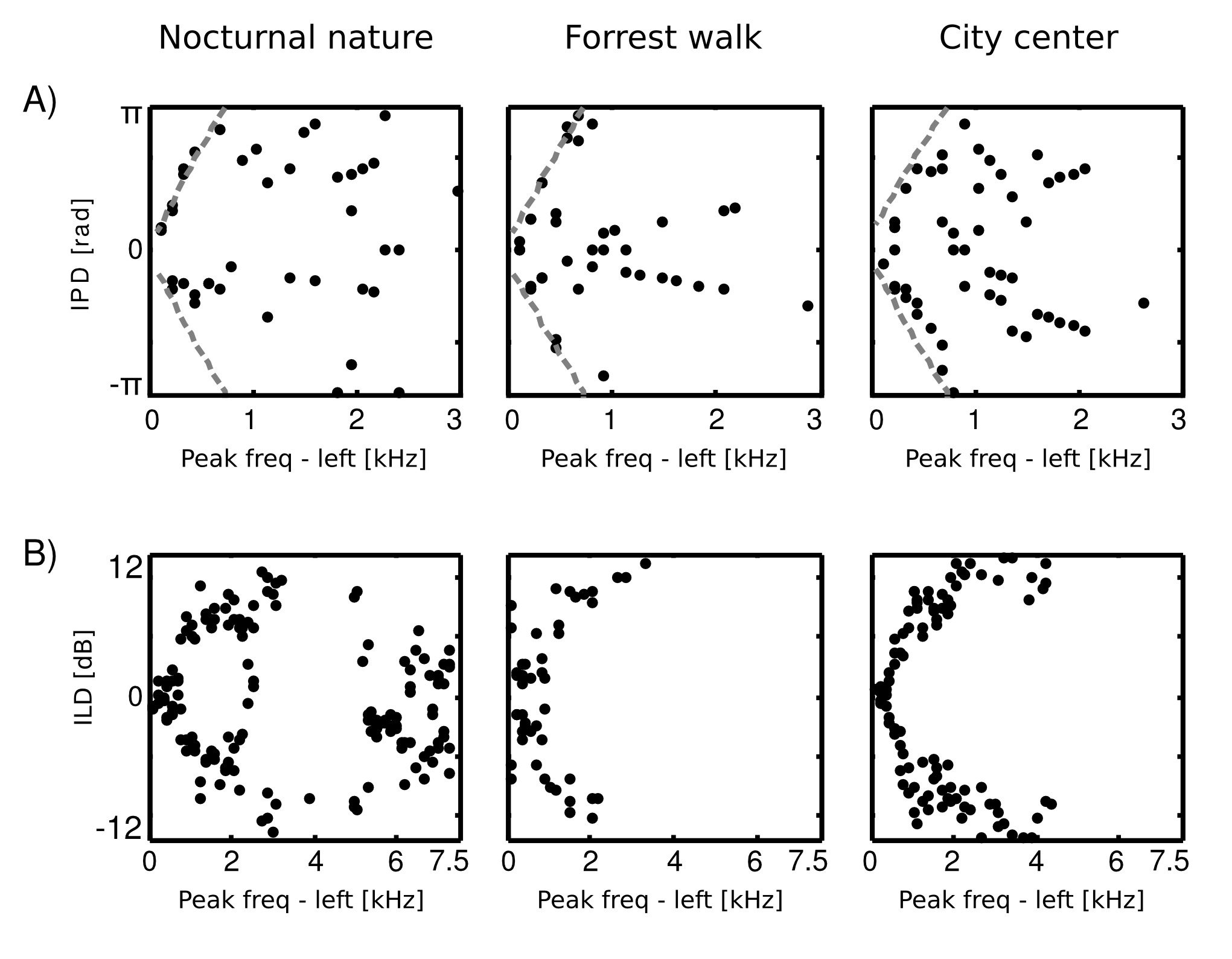}
\caption{Binaural cues represented by ICs capturing the same frequency in each ear. A) IPD as a function of frequency B) ILD as a function of frequency.}
\label{fig:ICA5}
\end{figure}

Prior to the analysis, raw recordings were down-sampled to the $22050$ Hz sampling rate.
The filtering and cue extraction pipeline is schematically depicted in figure \ref{schema}

To emulate spectral decomposition of the signal performed by the cochlea, sound waveforms from each ear were transformed using a filterbank of $64$ linear gammatone filters. Filter center frequencies were lineary spaced between $200$ and $3000$ Hz for IPD analysis and $200$ and $10000$ Hz for ILD analysis. 

A Hilbert transform in each frequency channel was performed. In result, instantenous phase $\phi_{L,R} (\omega, t)$ and amplitude $A_{L,R}(\omega, t)$ were extracted, separating level and phase information. Instantenous binaural cue values were computed in corresponding frequency channels $\omega$ from both ears according to the following equations:
\begin{equation}
\label{ILD}
ILD(\omega, t) = 10 \times \log_{10} \frac{A_L(\omega, t)}{A_R(\omega, t)}
\end{equation}

\begin{equation}
\label{IPD}
IPD(\omega, t)=\phi_L(\omega, t) - \phi_{R}(\omega, t) 
\end{equation}
IPDs with an absolute value exceeding $\Pi$ were wrapped to a $[-\Pi, \Pi]$ interval. 
Time series of IPD and ILD cues obtained in this way in each frequency channel were subjected to the further analysis.

\subsection*{Independent Component Analysis}

Independent Component Analysis (ICA) is a family of algorithms which
attempt to find a linear transformation of the data that minimizes redundancy \cite{HyvBook}. Given the data matrix $X \in \mathbb{R}^{n  \times m}$ (where $n$ is the number of data dimensions and $m$
number of samples), ICA finds a filter matrix $W \in \mathbb{R}^{n
  \times n}$ with
\begin{equation}
WX=S
\end{equation}
where the columns of $X$ are data vectors $x \in \mathbb{R}^n$, the  rows of $W$ are linear filters $w \in \mathbb{R}^n$ and $S \in \mathbb{R}^{n\times m}$ is a matrix of latent coefficients, which according to the assumptions are marginally independent. 
Equivalently the model can be defined using a basis function matrix $A=W^{-1}$, such that:
\begin{equation}
\label{bfMatrix}
X=AS
\end{equation}
The columns $a \in \mathbb{R}^n$ of  the matrix $A$ are called basis functions. In modelling of neural systems they are usually interpreted as linear receptive fields forming an efficient code of the training data ensemble \cite{HyvBook}. Each data vector can be represented as a linear combination of basis functions $a$, multiplied by linear coefficients $s$ according to the equation \ref{ICAsum}.
\begin{equation}
\label{ICAsum}
x(t)=\sum_i s_i a_i(t)
\end{equation}
where $t$ indexes the data dimensions. The set of basis functions $a$ is called a dictionary.
ICA attempts to learn a linear, maximally non-redundant code, hence the latent coefficients $s$ are assumed to be statistically independent i.e.
\begin{equation}
\displaystyle
p(s)=\prod_{i=1}^n p(s_i)
\end{equation}
The marginal probability distributions $p(s_i)$ are typically assumed to be sparse (i.e. of high kurtosis), since natural sounds and images have an intrinsically sparse structure \cite{OlshausenV1} and can be represented as a combination of a small number of primitives. In the current work we assumed a logistic distribution of the form:
\begin{equation}
p(s_i|\mu,\xi)=\frac{\exp(-\frac{s_i-\mu}{\xi})}{\xi(1+\exp(-\frac{s_i-\mu}{\xi}))^2}
\end{equation}
with position $\mu=0$ and the scale parameter $\xi=1$. 
Basis functions were learned by maximizing the log-likelihood of the model via gradient ascent \cite{HyvBook}.

Prior to ICA learning, the recordings were downsampled to a $14700$ Hz sampling rate (to obtain easy comparison with results in \cite{Lewicki}). A training dataset was created by randomly drawing $100000$ intervals each $128$ samples long (corresponding to $8.7$ ms). 

\section*{Acknowledgments}
This work was funded by the DFG graduate college InterNeuro.

\bibliographystyle{plain}

\section*{Figures}

\end{document}